# Improving knowledge of the acoustic factors involved in railway noise annoyance: first results of a pilot field survey of sixty-two local residents

Matthieu Sineau[1], Manuel Hellot, Robin Mafféïs,
Fanny Mietlicki, Laure Motio, Sarah Cherfan, Julien Bassinot
Bruitparif
32 Boulevard Ornano, 93200 Saint-Denis, France

Anne-Sophie Evrard[2]
Université Gustave Eiffel, Université Lyon 1, Umrestte, UMR T_9405
25, avenue François Mitterrand, Case24
F-69675 Bron Cedex • FRANCE Bron, France

Jean-Philippe Regairaz[3], Christophe Rosin
SNCF-Réseau
Direction Générale Industrielle et Ingénierie, Direction Technique du Réseau
15/17 rue Jean-Philippe Rameau
93210 La Plaine Saint-Denis, France

## ABSTRACT

*Railway transportation contributes to the objectives of decarbonization but also generates negative externalities, including noise. Energy noise indicators used to characterize population exposure do not adequately reflect the repetitive nature of railway noise peaks. The GENIFER pilot study aims to test a protocol designed to characterize railway noise events according to the instantaneous perceived annoyance when the train is passing, in order to improve understanding of the influence of acoustic factors on annoyance. The first phase of the survey was carried out in 2023 among 62 residents of a pilot site. An electronic device was used to collect around 5,000 ratings, ranging from 1 to 10, assessing the instantaneous annoyance induced by railway noise at passing trains. The site instrumentation included sixteen sound level meters and two video recording systems, enabling annoyance ratings to be associated with the acoustic characteristics of railway noise events. A questionnaire aimed at identifying co-determinants of long-term annoyance was also administered to participants. Feedback on the field implementation of this survey and initial results concerning acoustic measurements, instantaneous annoyance ratings and questionnaire responses will be presented.*

[1] matthieu.sineau@bruitparif.fr
[2] anne-sophie.evrard@univ-eiffel.fr
[3] jean-philippe.regairaz@reseau.sncf.fr

1. INTRODUCTION

Railway transportation has many environmental advantages in terms of decarbonizing the economy and travel is an appropriate response to the challenges of mass mobility. However, the noise generated by railway traffic represents a major negative externality of this mode of transportation, which can sometimes even be a brake on its development. The health impacts of railway noise, even if it affects fewer people than road traffic noise, are now well established scientifically, particularly in terms of long-term annoyance and sleep disturbance, and have been strongly recommended by the WHO [1]. However, it appears that the instantaneous annoyance due to railway noise is less documented. The discussions conducted in France by the Ministry of Ecological Transition and the French National Noise Council on these subjects have revealed the interest of introducing event-based indicators and railway event counting into the regulations, in addition to energy indicators [2].

The GENIFER project [3] (improving knowledge of the acoustic factors of the instantaneous annoyance due to railway noise), implemented by Bruitparif, the Gustave Eiffel University and SNCF Réseau, is part of this framework and proposes to carry out a feasibility study seeking to better understand the acoustic factors involved in the instantaneous annoyance expressed by residents regarding noise generated by railway traffic. It aims to categorise and rank railway noise events according to the level of instantaneous annoyance caused to railway neighbours and should lead to a better comprehension of the role that different acoustic characteristics of railway noise peaks may play in this annoyance. It will allow us to assess the relevance of conducting a larger study on a national scale.

Different methods were implemented and tested in this feasibility study: semi-structured interviews with 10 people, questionnaires and rating of instantaneous annoyance with about 60 participants, and commented listening to train sound samples recorded at the pilot site with 30 participants. This paper presents the first results of exploratory analysis of the instantaneous annoyance ratings in relation with the acoustic measurements conducted to characterise railway traffic, taking into account some information about the participants collected in the questionnaire.

2. MATERIAL AND METHODS

2.1. Selection of the pilot site and participants

The GENIFER study was carried out on a pilot site located in the commune of Savigny-sur-Orge (Figure 1) in the Ile-de-France region. It was selected based on different criteria in terms of rail traffic and populations according to their railway noise exposure [3].

The target was to find 20 people whose homes are exposed to more than 73 dB(A) in Lden [4], 20 people exposed to between 63 and 73 dB(A) and 20 people exposed to between 54 [1] and 63 dB(A). The selection criteria for the participants were as follows: the exposure of the dwelling to railway noise should be within the groups defined above, they should be at least 18 years old (legal age at the time of the survey), they should have no significant hearing problems and they should have lived in the dwelling for at least 6 months. Only one participant was selected per dwelling. Participants were recruited directly by Bruitparif using flyers distribution and door-to-door canvassing.

In total, 53 adults (25 males, 28 females; mean age = 50 ± 16 (SD)) participated in this phase of the study.

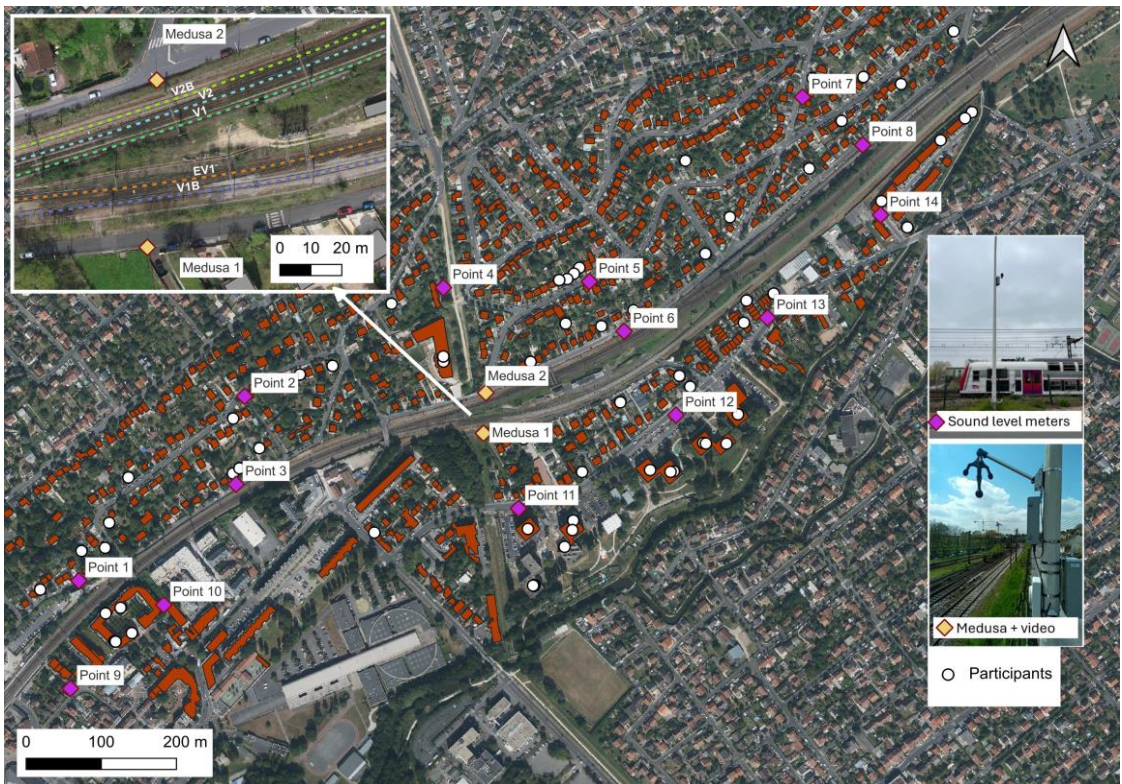

**Figure 1: View of the pilot site and the instrumental set-up**

## 2.2. Railway traffic

Characterizing train traffic was essential for identifying train observed by each participant. Based on train traffic data provided by France's national railway infrastructure manager (SNCF-RESEAU), traffic at the pilot site was estimated at around 350 trains per day. Throughout the study, we categorized train passing-by in five distinct groups: urban passenger trains *(RER)*, old generation regional trains *(CORAIL)*, new generation regional trains *(TER_NG2N)*, new generation regional short trains *(TER_AUTORAIL)*, and freights *(FRET)*. Preferential tracks are shown in Figure 1.

Two wide-angle cameras with a frame rate of 25 fps were installed on both sides of the tracks for train characterization with rail zone detecting movement recording. Image-based algorithms utilizing the camera's videos were implemented to know the track and the speed of the trains in the center of the site.

## 2.3. Acoustic measurements

To assess the railway noise exposure of the participants during the survey, 16 noise measurement systems were installed all over the study area during the survey: 14 stand-alone sound level meters spread throughout the study area and two expert systems (Medusa) in the centre of the study area on each side of the tracks (Figure 1). The Medusa system has been developed by Bruitparif and has a functionality that detects the direction of the sound. It is thus possible to detect all sound coming from rails in an exhaustive manner.

These measurement systems provide Leq,100 milliseconds for A and C weighted levels. The analysis of the sound level meters was conducted through automatic detection of railway events using a low-pass filter empirical method implemented by Bruitparif for its railway noise monitoring network. This method, employing a Butterworth filter, involves filtering the $L_{Aeq100ms}$ with two different degrees of low-pass filters (n=5 and n=1) and cut-off frequencies (Wn=0.02 and Wn=0.002). The intersection of the two filtered signals marks the beginning of the acoustic event, while their subsequent intersection signifies the event's end. In absence of a standardized method, for railway noise monitoring systems, to accurately determine start/end events timestamps, the intersections of the two filtered signals provides a good estimation of the total event duration, including in part the approach of the train and its distance. This duration estimation seems more realistic than an estimation based on train time pass-by (as $L_{Aeq,tp}$) or duration within the "LAmax - 10dB" range, as is sometimes performed [5]. Thresholds based on event duration, emergence, and minimum $L_{Aeq}$ level were then applied to classify them as railway events (Figure 2). The expert systems allowed us to add another layer of verification by using the localization of the sound. Due to the challenges associated with detecting train events in densely urbanized areas, only the dataset from the Medusas was used. The sound level meters data were used to validate the numerical modelling. The acoustic descriptors calculated for each railway noise event are those listed by the French National Noise Council [2][5].

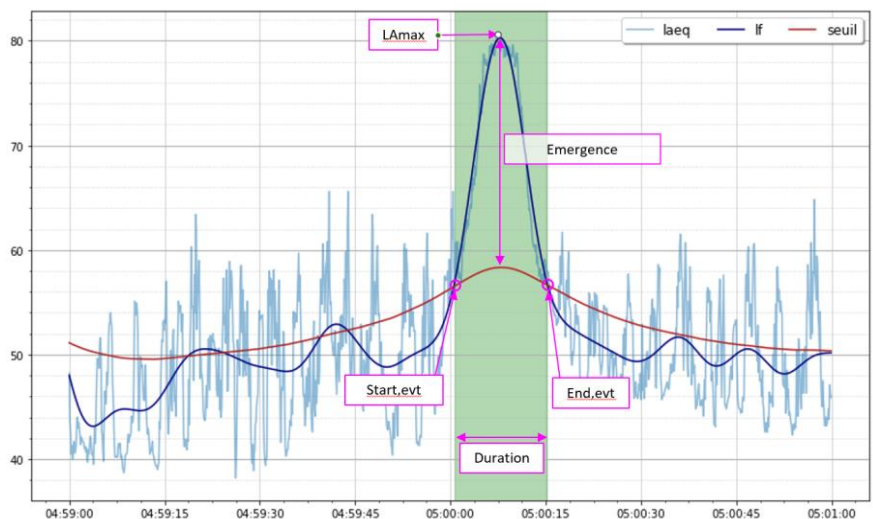

**Figure 2: Detection of railway events with low-pass filter method**

A numerical modelling of railway noise was carried out with CadnAa® on the entire pilot site to simulate the sound levels generated by the passage of a train at any point in the territory and to determine the corrections to be applied to the acoustic descriptors measured at the measuring expert stations (Medusa). An evaluation of the acoustic descriptors at the facades of the dwellings of each participant have been obtained. Railway noise indicators associated with the instantaneous annoyance ratings have been corrected with the offsets determined by the numerical modelling.

**2.4. Instantaneous annoyance ratings**

A central part of the survey was for participants to record their level of instantaneous annoyance when trains of different categories pass by. In order to do this, they had to use a remote control ("Noisemote", Figure 3). This remote control was connected to the Bruitparif server to determine the precise time of the train's passage. Each participant had to carry out a total of three hours of train rating in six sessions of 30 minutes or three sessions of one hour. The sessions were dedicated exclusively to the rating of trains and not to other activities. All the instantaneous annoyance ratings, on a scale of 1, lowest annoyance, to 10, highest annoyance, were stored in a database. Instantaneous annoyance ratings were mainly given during the daytime.

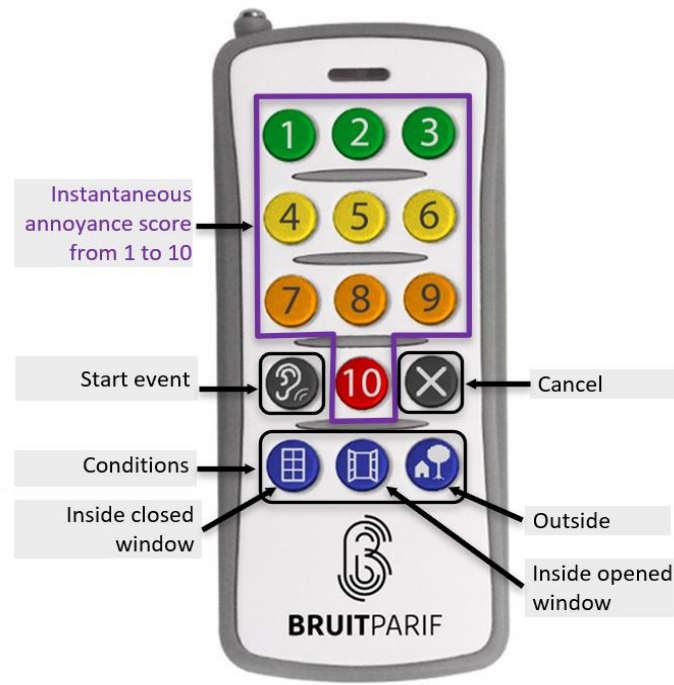

Figure 3: The remote control "NOISEMOTE" used in the experiment

## 2.5. Statistical methods

For each instantaneous annoyance rating, the start of the event, the conditions of notation (opened window, closed window, outside), information about the participants collected in the questionnaire (railway noise exposure zone of their dwelling, windows type in terms of acoustic insulation, long term high annoyance (HA) due to all sources of noise to which participants were exposed (called global noise), long term HA due to railway noise, Weinstein noise sensitivity score [6]), and railway noise indicators (SEL, $L_{Amax}$, $L_{Ceq}$-$L_{Aeq}$, duration) were stored in the database. Long term global noise and railway noise annoyance were assessed using the ISO/ICBEN (International Commission on the Biological Effects of Noise) recommended question [7]: "Thinking about the last 12 months when you are here at home, how much does global noise / railway noise bother, disturb or annoy you?" The standard verbal scale was used with five possible answers: extremely, very, moderately, slightly or not at all. High annoyance was defined by the proportion of highly annoyed people (%HA), i.e. by the proportion of people reporting to be very or extremely annoyed by global or railway noise [8].

The acoustic data corresponding to each instantaneous annoyance rating were incorporated into the database using the date of the events (nearest temporal join) and this incorporation was checked manually.

In order to make preliminary use of the data, the instantaneous annoyance ratings were assumed to be independent and identically distributed.

Because of the presence of quantitative and qualitative variables, a factorial analysis of mixed data (FAMD) was used to study the proximity of variables to each other and to observations [8]. The analyses were performed on R software using "FactoMineR" [9].

Based on the results of the factorial analysis, an ascending hierarchical classification [10] was performed to assess the relevance of using clusters and to have a better visualisation of the data [11]. This hierarchical classification was carried out to categorise the different types of trains according to the instantaneous annoyance they cause, their acoustic characteristics and non-acoustic factors potentially involved in the annoyance reported by the participants. Classification was performed using Ward's method and Euclidean distance as the metric.

A chi-2 test was applied to the categorical variables and an ANOVA to the quantitative variables to assess the overall relationship between these variables and the cluster. Cluster by cluster analysis of the V-Test values [12] showed the differences in the composition of each group by comparing the averages of the inter-cluster variables with the average for the same variable in the full sample.

## 3. RESULTS

### 3.1. Descriptive analysis of railway traffic

Table 1 presents the descriptive analysis of the railway traffic during the study. Acoustic measurements were recorded with the medusa 2. Nearly 63 % of the traffic consisted of RER trains, 17% freights, 13% CORAIL trains, 6% TER_NG2N trains, and 1% TER_AUTORAIL trains.

Freights had the lowest average speed (56 km/h) whereas regional trains had the highest at 130 km/h. The noisiest trains were CORAIL trains with an average SEL of 99.5 dB(A), followed by freight, TER NG2N, RER without stops, RER with stops, and TER AUTORAIL (Table 1).

**Table 1: Descriptive analysis of railway traffic over the entire survey period**

| Train groups | Type of train | % of traffic | Speed km/h | SEL dB(A) | Preferential tracks |
|---|---|---|---|---|---|
| CORAIL | old generation regional trains | 13% | 130.8 ± 24.7 | 99.5 ± 4.9 | V1, V2 |
| FRET | Freight | 17% | 57.3 ± 24.4 | 92.7 ± 5.9 | V2B, EV1 |
| RER with stops | urban passenger trains | 56% | 78.2 ± 17.3 | 85.7 ± 4.4 | V1B, V2B |
| RER without stops | urban passenger trains | 7% | 114.7 ± 16.8 | 87.1 ± 5.2 | V1, V2 |
| TER_AUTORAIL | new generation regional short trains | 1% | 125.2 ± 26.3 | 85.3 ± 5.6 | V1, V2 |
| TER_NG2N | new generation regional trains | 6% | 134.3 ± 26.1 | 91.4 ± 5.4 | V1, V2 |

## 3.2. Hierarchical clustering

The calculated tree suggested limiting the number of clusters to three (Figure 4) to obtain sufficient inertial gain [10].

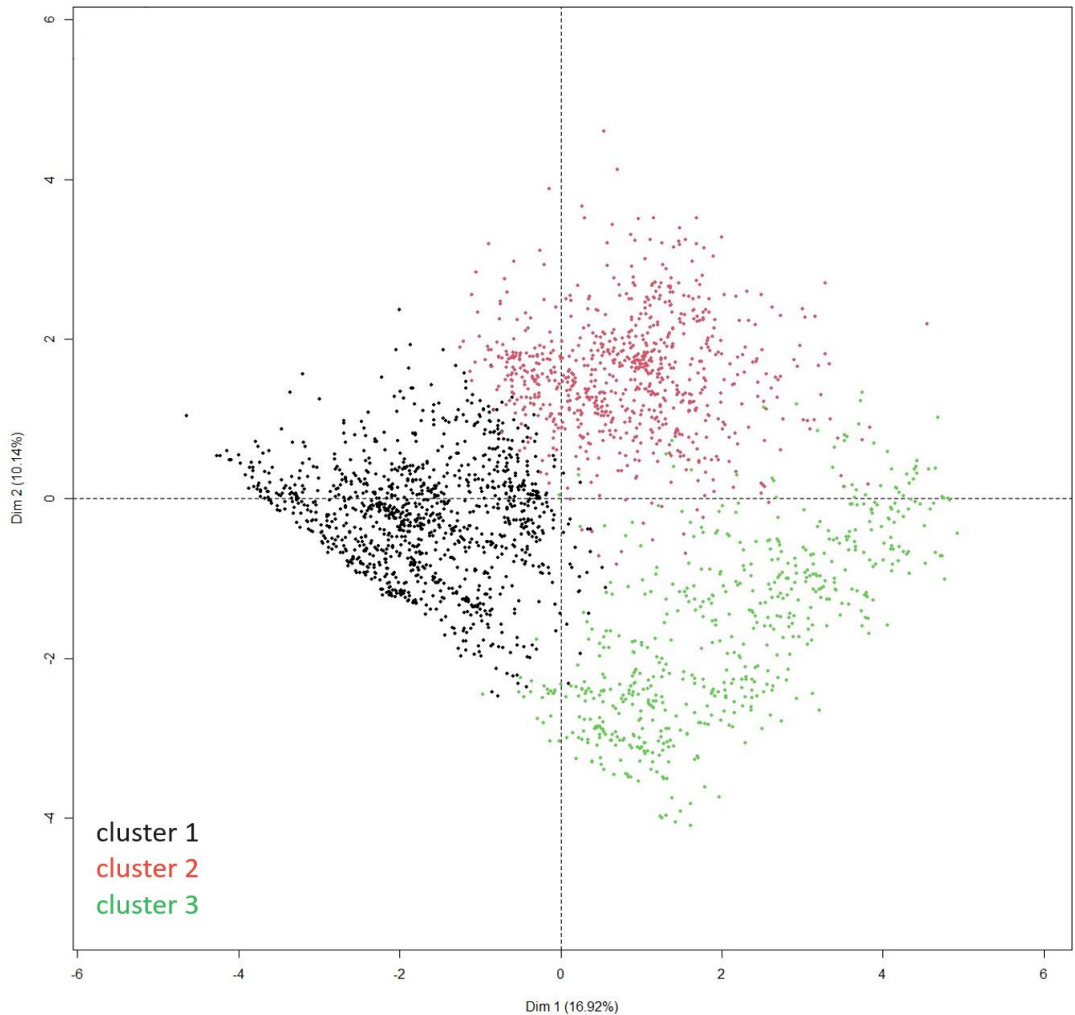

**Figure 4: Hierarchical classification on the first two principal components of the FAMD**

The hierarchical classification by principal components revealed three clusters (Table 2):
- A group (cluster 1) with lower average instantaneous annoyance ratings (mean = 4.5), a high proportion of instantaneous annoyance ratings associated with RER trains (87 %) and non-highly annoyed individuals (85 % for long-term annoyance due to railway noise and 97 % for long-term annoyance due to global noise).
- A group (cluster 2) with lower average instantaneous annoyance ratings (mean = 4.5), a high proportion of instantaneous annoyance ratings associated with RER trains (83 %) and with greater representation of highly annoyed participants (98 % for long-term annoyance due to railway noise and 43 % for long-term annoyance due to global noise).
- A group (cluster 3) with higher average instantaneous annoyance ratings (mean = 6.2) than the overall average (mean = 4.9), a high proportion of Corail (90%), and higher noise levels (mean SEL_calibrated = 91 dB(A)), but with no statistically significant (p-value > 0.05) difference between HA and non-HA people (for both long-term annoyance due to railway noise or global noise).

Freight trains did not appear much in the clusters because they were rarely rated by participants because of mainly night-time passages. Freight trains represented 4.2 % of the total number of instantaneous annoyance ratings, even though these trains accounted for 17 % of overall railway traffic on the site. In contrast, Corails were over-represented, 21 % among the total number of instantaneous annoyance ratings, even though these trains accounted for 13 % of overall rail traffic. Other trains have been rated in the same proportions to those observed over the entire study period.

**Table 2: Description of each cluster with qualitative and quantitative variables**

| Variable | Type | Cluster 1 (n=1150) | Cluster 2 (n=864) | Cluster 3 (n=553) | Global |
|---|---|---|---|---|---|
| | | Value | Value | Value | Value |
| Instantaneous annoyance ratings | Mean ± SD | 4.5 ± 2.3 | 4.5 ± 2.2 | 6.2 ± 2.2 | 4.9 ± 2.4 |
| SEL_{recalibrated} | Mean dB(A) ± SD | 71.7 ± 8.1 | 81.8 ± 6.3 | 91.1 ± 7.6 | 79.5 ± 10.8 |
| RER | Qualitative** | 87.3% (59.5%) | 82.5% (40.3%) | 0.5% (0.2%) | 65.4 % |
| CORAIL | Qualitative** | 0.4% (0.9%) | 0.2% (0.4%) | 89.8 % (98.7%) | 21.1 % |
| FRET | Qualitative** | 2.6% (27.8%) | N/A* (37%) | 6.4% (35.2%) | 4.2 % |
| TER NG2N | Qualitative** | N/A* (40.1%) | 8.3% (46.3%) | 3.4% (13.6%) | 5.7 % |
| TER AUTORAIL | Qualitative** | 3.7% (60.9%) | N/A* (39.1%) | 0% (0%) | 2.7 % |
| HA_train_noise | Qualitative** | 15.0 % (13.5 %) | 97.8% (63%) | N/A* (23.5%) | 49.7 % |
| HA_global_noise | Qualitative** | 2.4% (5.6%) | 43.2% (70.7%) | N/A* (23.7%) | 19.6 % |

*N/A when P-value >0.05*
***Qualitative variables are presented in the following format: X%(Y%), where X% represents the percentage of number of variable samples per total samples in the cluster and Y% represents the percentage of variable samples within the cluster per total number of variable samples.*

## 4. DISCUSSION AND CONCLUSION

This feasibility study made it possible to assess the instantaneous annoyance rating, caused by trains pass-by, using a remote control under different conditions. It also made it possible to assess the relevance of the protocol for ranking different types of train according to their characteristics and to the associated instantaneous annoyance. The scoring of instantaneous annoyance using the Noisemote was well accepted by the participants. Except for nine dropouts (less than 7%), participants all agreed to spend at least a total of three hours rating trains annoyance.

Clustering seems to indicate that for the noisiest trains, above certain thresholds yet to be determined in terms of level (for SEL or $L_{max}$), all people tended to give higher instantaneous annoyance ratings whether they were highly annoyed (in terms of long-term annoyance) by the noise or not. The statistical analysis of the data using mixed models [13] would enable the exploration of relationships between the dependency of annoyance ratings on each other and the effects of groups of variables.

In this study, the individual noise exposure of each participant was not measured by a sound level meter. Because of that, to characterise the railway noise of each participant has been a complicated task. Given the difficulty of detecting railway noise events by measurements spread over the study area, it was decided to use precise and exhaustive data from Medusa systems near the tracks and then to apply a sound propagation offset calculated track by track from a numerical modelling. This method has the advantage of being based on an accurate detection of railway noise events but has the disadvantage of a potentially large uncertainty, particularly for the receptors furthest from the tracks. In addition, railway noise levels inside dwellings were not assessed.

In a large-scale study, it would seem necessary to instrument all the participants with individual sound levels meters to measure their noise exposure levels accurately, whatever the conditions (outside / inside the dwelling).

Additionally, to enable a direct comparison of instantaneous annoyance ratings across different individuals while mitigating the influence of participants providing more ratings than others, it would be preferable to establish common rating periods under identical conditions for all participants.

A large-scale study should also include assessment of medium-term annoyance (e.g. daily) to establish the relationship between instantaneous annoyance and long-term annoyance. This would provide a better understanding of the influence of the number of noise events on annoyance according to their acoustic characteristics and period of occurrence. This approach would pave the way for the development of a noise points counter.

**FUNDING**

This project was supported by the French National Research Program for Environmental and Occupational Health (ANSES-22-EST-182) of ANSES (French Agency for Food, Environmental and Occupational Health & Safety).

**ACKNOWLEDGEMENTS**

The authors would like to thank the ANSES for its assistance. They thank Chrystèle Philipps-Bertin and Patricia Champelovier of the AME-MODIS department of the Gustave Eiffel University for their help in developing the interview grid and the questionnaire. They would also like to thank the Municipality of Savigny-sur-Orge and the Grand Orly Seine-Bièvre public territorial establishment for their help in implementing this study on their territory. Finally, the authors would like to thank the inhabitants of Savigny-sur-Orge for their active participation and involvement in this survey.